\documentstyle[preprint,prc,aps,floats,epsf]{revtex}
\begin{document}
\title{Nucleons or diquarks? Competition between clustering and color 
superconductivity in quark matter}
\author{St\'ephane Pepin\footnote{Current address: Max-Planck Institut f\"ur 
Kernphysik, Postfach 103980, D-69029 Heidelberg, Germany}, 
Michael C. Birse and Judith A. McGovern}
\address{Theoretical Physics Group,
Department of Physics and Astronomy, University of Manchester, Manchester
 M13 9PL, United Kingdom}
\author{Niels R. Walet}
\address{Department of Physics, UMIST, P.O. Box 88, Manchester M60 1QD,
United Kingdom}
\date{\today}
\maketitle
\begin{abstract}
We study the instabilities of quark matter in the framework of a generalized 
Nambu--Jona-Lasinio model, in order to explore possible competition between 
three-quark clustering to form nucleons and diquark formation leading to color 
superconductivity. Nucleon and $\Delta$ solutions are obtained for the 
relativistic Faddeev equation at finite density and their binding energies 
are compared with those for the scalar and axial-vector diquarks found from 
the Bethe-Salpeter equation. In a model with interactions in both scalar
and axial diquark channels, bound nucleons exist up to nuclear matter density.
However, except at densities below about a quarter of that of nuclear matter,
we find that scalar diquark formation is energetically favored. This raises the 
question of whether a realistic phase diagram of baryonic matter can be 
obtained from any model which does not incorporate color confinement.
\end{abstract}

\section{INTRODUCTION}
In contrast to QCD at finite temperature, rather little is known about 
QCD at finite density. Technical difficulties (such as the complex fermionic 
determinant at finite chemical potential) make lattice 
Monte-Carlo simulations very difficult. Even though some techniques are being
developed to overcome these problems (for example, the Glasgow 
method\cite{BMK98} or the technique of imaginary chemical 
potential\cite{AKW99}), they are not yet able to provide unambiguous results.

However, models of QCD seem to indicate a rich phase structure in high-density 
quark matter. In particular, much attention has recently been devoted to 
so-called ``color superconductivity"\cite{ARW98,RSSV98}: an arbitrarily weak 
attractive force makes the Fermi sea of quarks unstable at high
density with respect to diquark formation and induces 
Cooper pairing (diquark condensation). Although this phenomenon 
had been studied earlier\cite{BL84,II95,DFL96}, the large magnitude of the 
superconducting gap found in the more recent studies\cite{ARW98,RSSV98} (more 
than 100 MeV) suggested that this was much more important than had been thought
previously and has generated an extensive literature. In 
Refs.~\cite{ARW98,RSSV98} an instanton model was used to calculate the gap at
finite density and zero temperature. Berges and Rajagopal \cite{BR99} extended
this work and calculated the phase diagram of strongly interacting matter as
a function of temperature and baryon-number density in the same model. 
Identical results were found in the Nambu--Jona-Lasinio model \cite{SKP99}.
The existence of such a color-superconducting gap could have important
consequences for the physics of neutron stars or even for heavy ion 
collisions\cite{ARWb98,Raja99}. (For a review of the field, see:\cite{Sch99}.) 

Previous studies of color superconductivity have focused on instabilities of 
the quark Fermi sea with respect to diquarks only. Of course, it is known that
at lower densities (of the order of nuclear matter density) three-quark
clusters---nucleons---are the dominant degrees of freedom. Here we
address the question of the possibility of a competition between diquark 
condensation and three-quark clustering at finite density. 
To answer such a question fully would require a three-particle generalization 
of the BCS treatment, a complicated task. As a first step towards this goal,
we look for instabilities of the quark Fermi sea with respect to three-quark 
clustering by studying the nucleon binding energy in quark matter with finite
density. A bound nucleon at finite density would be a signal of instability 
with respect to clustering (in the same way that a bound diquark at finite 
density is a signal of instability with respect to diquark condensation). A
comparison of the magnitudes of the binding energies of the diquark and the 
nucleon can give some idea of the relevance of these degrees of freedom.
Very recently, Beyer {\it et al.} have studied a similar clustering
problem for three nucleons in nuclear matter\cite{BSKR99}.

To perform this study we solve the Bethe-Salpeter equation for the diquarks
and the Faddeev equation for the nucleon (and also, for completeness, for the
$\Delta$) within the framework of a Nambu--Jona-Lasinio (NJL) model. The use of 
a separable interaction simplifies the treatment of the Faddeev equation
considerably. The form of the equation reduces to that of a Bethe-Salpeter 
equation describing the interaction between a quark and a diquark. At zero 
density, several groups have used this Faddeev approach to study baryons in 
the NJL model (for example, Refs.~\cite{BAR92,IBY93,Me94,HT94,IBY95,HK95}). 
More recent papers\cite{Oet97} have extended this treatment to incorporate a
mechanism for confinement. In this work we will mainly follow the formalism 
developed by Ishii, Bentz and Yazaki\cite{IBY95}, generalizing it to finite 
density. In this type of approach, it is crucial to include the axial-vector 
diquark channel in addition to the scalar channel, as otherwise the nucleon 
is very weakly bound. In previous preliminary studies, keeping only the scalar
channel, we have found that binding of a nucleon in matter is only possible at 
very low densities, less than 10\% of nuclear matter density\cite{BMPW99}.

The paper is organized as follows: in the next section, we briefly describe 
the NJL model and its application to quark-quark interaction. The parameters 
used in our study are also discussed. In Sec.~3 the Bethe-Salpeter equations
in the scalar and axial-vector diquark channels are solved. The form of the 
Faddeev equation for the nucleon is presented in Sec.~4. In Sec.~5, we describe 
the numerical techniques used and in Sec.~6 we present our results for the 
nucleon and the $\Delta$ at finite density. Finally, we draw some conclusions 
in Sec.~7.
  
\section{THE MODEL}
The NJL model provides a simple implementation of dynamically broken chiral 
symmetry, based on a two-body contact interaction\cite{NJL61}. In spite of the 
fact that the model does not incorporate confinement, it has been successfully 
applied to the description of mesonic properties at low energy. (For reviews, 
see Refs.~\cite{Kle92,HK94}.)
 
The model Lagrangian has the form
\begin{equation}
{\cal L}=\overline\psi(i\partial\llap/-m)\psi+{\cal L}_I,
\end{equation}
where ${\cal L}_I$ is the interaction Lagrangian. In the present work we 
consider only the chiral limit, setting the current quark mass $m$ to zero.
Several versions of the NJL interaction can be found in the literature.
The original version\cite{NJL61} is
\begin{equation}\label{orig}
{\cal L}_{I} =  g \Bigl[ (\overline{\psi}\psi)^2 + (\overline{\psi}
i\gamma_5\vec{\tau}\psi)^2 \Bigr].
\end{equation}
One can also work with a color-current interaction, 
\begin{equation}\label{OGE}
{\cal L}_{I} = -g \sum_c \left(\overline{\psi}\gamma_\mu
{\textstyle\frac{1}{2}}\lambda_c\psi\right)^2,
\end{equation}
where $\lambda_c$ ($c=1,...,8$) are the usual Gell-Mann matrices. 

Whatever version of the model is chosen, a Fierz transformation should
be performed in order to antisymmetrize the interaction
Lagrangian. This allows ${\cal L}_I$ to be brought into a form where
the interaction strength in a particular channel can be read off
directly from its coefficient in the Lagrangian.  For the
$q\overline{q}$ channel, one just rewrites ${\cal L}_I$ into the form
${\cal L}_{I,q\overline{q}} = \frac{1}{2} ({\cal L}_I + {\cal
L}_{I,F})$ where ${\cal L}_{I,F}$ is the Fierz transformed form of
${\cal L}_I$. We shall need here only the scalar and pseudoscalar
terms of the $q\overline{q}$ interaction.  These have the same form as
Eq.~(\ref{orig}), but with a coupling constant $g_{\pi}$ which is
related to the original coupling constant $g$ of the Lagrangian by a
coefficient given by the Fierz transformation. For example, one has
$g_\pi/g = 13/12$ for the model defined by Eq.~(\ref{orig}) and
$g_\pi/g = -2/9$ for Eq.~(\ref{OGE}).

To study the nucleon we also need to rewrite the interaction
Lagrangian in the form of a $qq$ interaction. This is done by a Fierz
transformation to the $qq$ channels, which allows the interaction to
be expressed as a sum of terms of the form $(\overline{\psi} A
\overline{\psi}^T)(\psi^T B \psi)$, where the matrices $A$ and $B$ are
overall antisymmetric in Dirac, isospin and color indices.  (We use
the Dirac representation for the $\gamma$-matrices and follow the
conventions of Itzykson and Zuber\cite{IZ80}.)

As our three-quark state must be a color singlet, the diquark channels
of interest are color anti-triplet. For a local interaction the relevant 
channels are the scalar ($0^+, T=0$) and axial-vector ($1^+,T=1$) ones. These 
are also the channels in which more realistic interactions (including, for 
example, one-gluon exchange) are expected to be most attractive. Explicitly, 
we have
\begin{equation}\label{scal}
{\cal L}_{\scriptscriptstyle S} = g_{\scriptscriptstyle S} 
\sum_a\left(\bar{\psi}\gamma_5 C\tau_2\beta^a\bar{\psi}^T\right)
\left(\psi^T C^{-1}\gamma_5\tau_2\beta^a\psi\right),
\end{equation}
for the scalar channel and
\begin{equation}\label{ax}
{\cal L}_{\scriptscriptstyle A} = g_{\scriptscriptstyle A} 
\sum_{i,a}\left(\bar{\psi}\gamma_\mu C\tau_i\tau_2\beta^a\bar{\psi}^T
\right)\left(\psi^T C^{-1}\gamma^\mu\tau_2\tau_i\beta^a\psi\right),
\end{equation}
for the axial-vector one. The matrices $\beta^a=\sqrt{3\over 2} \, \lambda^a$ 
for $a=2,5,7$ project onto the color 
$\bar{3}$ channel and $C=i\gamma_2\gamma_0$ is the charge conjugation matrix.   
The coupling strengths $g_{\scriptscriptstyle S}$ and 
$g_{\scriptscriptstyle A}$ are again related to the original $g$ by 
a coefficient given by the Fierz transformation. In the following we do not
choose a specific version of the NJL Lagrangian but instead treat the physical 
couplings $g_\pi,g_{\scriptscriptstyle S},g_{\scriptscriptstyle A}$ as 
independent parameters.

The gap equation for the constituent quark mass $M$ reads
\begin{equation}\label{gap} 
M = 2 i g_\pi \int \frac{d^4k}{(2 \pi)^4} {\rm Tr}[S(k)],
\end{equation}
where the quark propagator is
\begin{equation}
S(k)  = \frac{1}{k\llap/ - M + i \epsilon}.
\end{equation}
The integral (\ref{gap}) diverges and so has to be regularized. There are 
various regularization schemes at our disposal: Pauli-Villars, proper-time,
and 3- or 4-momentum cut-off. In this work we use a sharp cut-off
$\Lambda$ on the 3-momentum to regularize the loop integrals, since
this is conveniently applied to systems of finite density.  Although a
3-momentum cut-off is not Lorentz invariant, this is less relevant at finite 
density where there is a much larger, physical, breaking of Lorentz invariance 
due to the presence of a quark Fermi sea. In any case, we shall show that
physical observables depend only weakly on the choice of regulator.

The two parameters $g_\pi$ and $\Lambda$ can be fitted to a given value of the
constituent quark mass and to the pion decay constant $f_\pi=93$ MeV. This last 
quantity is evaluated from
\begin{equation}
f_\pi^2 = -12 i M^2 \int \frac{d^4k}{(2 \pi)^4} \frac{1}{(p^2-M^2)^2}.
\end{equation}
In the following calculations we use two different values for the constituent 
mass $M=450$ MeV and $M=500$ MeV, both of which have been chosen to be large 
enough that the mass of the $\Delta$ lies below the three-quark threshold. In 
Table \ref{para} we list the values of $g_\pi$ and $\Lambda$ for the 
corresponding constituent masses. As one can see, the values for the cut-off 
are relatively low and decrease with increasing constituent mass. One cannot 
push the constituent mass to higher values as otherwise the cut-off would 
approach (or, worse, become smaller than) the quark mass.

\begin{table}
\caption[parameters]{\label{para} Values of the parameters $g_\pi$ and 
$\Lambda$ for the two values we use for the constituent quark mass $M$. 
For each $M$, the $qq$ coupling ratios $g_{\scriptscriptstyle S}/g_\pi$ and 
$g_{\scriptscriptstyle A}/g_\pi$
are determined by fitting the $N$ and $\Delta$ masses. In each case we also
list, in parentheses, the minimal value of the coupling ratio 
required to produce  bound diquarks.\\}
\begin{tabular}{ccccc}
 $M$ & $g_\pi$ & $\Lambda$ & $g_{\scriptscriptstyle S}/g_\pi$  (min.) & 
$g_{\scriptscriptstyle A}/g_\pi$  (min.)\\
(GeV) & (GeV$^{-2}$) & (GeV) \\
\tableline
0.45 & 7.914 & 0.579 & 0.72  (0.33) & 0.46  (0.37)\\
0.5  & 8.634 & 0.573 & 0.72  (0.28) & 0.53  (0.31)
\end{tabular}
\end{table}

\begin{figure}
\epsfysize=8cm \centerline{\epsffile{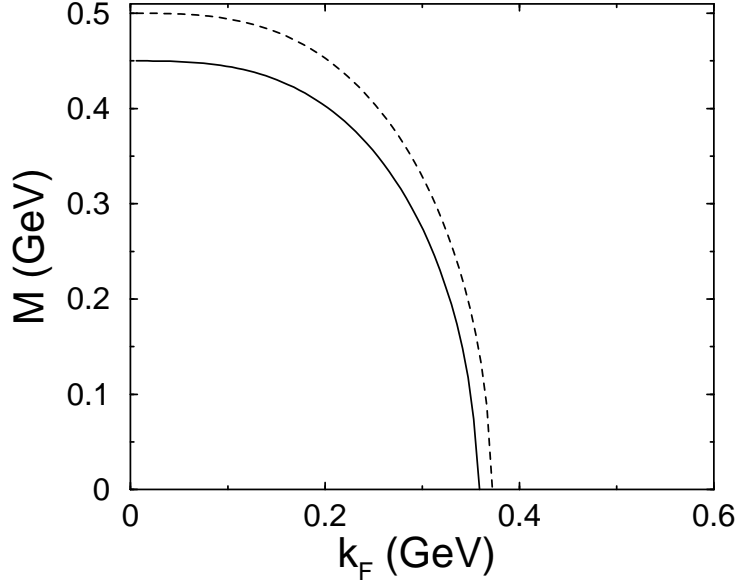}}
\caption{Evolution of the constituent quark mass $M$ with the Fermi 
momentum $k_F$ for $M(k_F=0)=450$ MeV (continuous curve) and 
$M(k_F=0)=500$ MeV (dashed).\label{mass}}
\end{figure}

The effects of finite density on the constituent mass are taken into account
by introducing the Fermi momentum $k_F$ as a lower cut-off on the integral
in the gap equation (\ref{gap}). In Figure \ref{mass} we show the evolution of 
the constituent mass as a function of the Fermi momentum for the two sets of 
parameters corresponding to  $M=450$ and 500 MeV. For these parameters, the 
restoration of chiral symmetry occurs at $k_F=359$ MeV and $k_F=372$ MeV 
respectively. (For comparison, nuclear matter density corresponds to a quark 
Fermi momentum of $k_F=[{3\over 2}\pi^2\rho_B]^{1/3}=270$ MeV).  

\section{THE TWO-BODY $t$-MATRIX}
The diquark $T$-matrix is an essential building block for the Faddeev equation. 
It is obtained by solving the Bethe-Salpeter equation in the ladder 
approximation,
\begin{equation}\label{BS}
T^{\alpha\beta,\gamma\delta}(k) = K^{\alpha\beta,\gamma\delta} +
{1\over 2} \sum_{\lambda,\epsilon,\lambda',\epsilon'}
\int \frac{d^4q}{(2 \pi)^4}K^{\alpha\beta,\lambda\epsilon} S^{\lambda
\lambda'}(k+q)S^{\epsilon\epsilon'}(-q)T^{\lambda'\epsilon',\gamma\delta}(k).
\end{equation}
In the scalar diquark channel, the interaction kernel $K$ is
\begin{equation}\label{kersc}
K_{\scriptscriptstyle S}^{\alpha\beta,\gamma\delta} 
= 4 i g_{\scriptscriptstyle S} \sum_a
(\gamma_5 C\tau_2\beta^a)^{\alpha\beta}
(C^{-1}\gamma_5\tau_2\beta^a)^{\gamma\delta}.
\end{equation}
This interaction is momentum-independent and so Eq.~(\ref{BS}) can be solved 
easily to get the diquark $T$-matrix in the scalar channel\cite{IBY95}:
\begin{equation}
T_{\scriptscriptstyle S}^{\alpha\beta,\gamma\delta}(k) 
= \sum_a (\gamma_5 C\tau_2\beta^a)^{\alpha\beta}
\bar{\tau}_{\scriptscriptstyle S}(k) 
(C^{-1}\gamma_5\tau_2\beta^a)^{\gamma\delta},
\end{equation}
with
\begin{equation}\label{tmasca}
\bar{\tau}_{\scriptscriptstyle S}(k) = 
\frac{4 i g_{\scriptscriptstyle S}}{1+2 g_{\scriptscriptstyle S} 
\Pi_{\scriptscriptstyle S}(k^2)},
\end{equation}
and
\begin{equation}
\Pi_{\scriptscriptstyle S}(k^2) = 6 i \int \frac{d^4q}{(2 \pi)^4} 
{\rm Tr}_D[\gamma_5 S(q) \gamma_5 S(k+q)].
\end{equation}
If the $T$-matrix (\ref{tmasca}) has a pole, this gives us the mass of the
bound scalar diquark. Note that, if one replaces the coupling 
$g_{\scriptscriptstyle S}$ by 
$g_\pi$, the denominator of (\ref{tmasca}) is the same as that in the pion 
$q\overline q$ channel\cite{IBY95}. This means that for 
$g_{\scriptscriptstyle S}=g_\pi$ the scalar 
diquark and pion are degenerate, and have zero mass in the chiral limit. Since
diquarks do not condense in the vacuum, this puts an upper limit to the choice 
of the scalar coupling $g_{\scriptscriptstyle S}$. 

In the axial-vector channel, the kernel is
\begin{equation}\label{kerax}
K_{\scriptscriptstyle A}^{\alpha\beta,\gamma\delta} 
= 4 i g_{\scriptscriptstyle A }\sum_{a,i}
(\gamma_\mu C\tau_i\tau_2\beta^a)^{\alpha\beta}
(C^{-1}\gamma^\mu\tau_2\tau_i\beta^a)^{\gamma\delta},
\end{equation}
and the solution of (\ref{BS}) can be shown to be
\begin{equation}
T_{\scriptscriptstyle A}^{\alpha\beta,\gamma\delta}(k) 
= \sum_{a,i} (\gamma_\mu C\tau_i\tau_2\beta^a)^{\alpha\beta} 
\bar{\tau}^{\mu\nu}_{\scriptscriptstyle A}(k) 
(C^{-1}\gamma_\nu\tau_2\tau_i\beta^a)^{\gamma\delta},
\end{equation}
with
\begin{equation}\label{tmaax}
\bar{\tau}^{\mu\nu}_{a}(k) = 4 i g_{\scriptscriptstyle A} 
\left[\frac{g^{\mu\nu} - k^\mu k^\nu/k^2}
{1 + 2 g_{\scriptscriptstyle A} \Pi_{\scriptscriptstyle A,T}(k^2)} 
+\frac{k^\mu k^\nu/k^2}{1 + 2 g_{\scriptscriptstyle A} 
\Pi_{\scriptscriptstyle A,L}(k^2)} \right].
\end{equation}
Here, the axial polarization tensor,
\begin{equation}
\Pi_{\scriptscriptstyle A}^{\mu\nu}(k)= 6 i \int \frac{d^4q}
{(2 \pi)^4} {\rm Tr}_D[\gamma^\mu S(q) \gamma^\nu S(k+q)],
\end{equation}
has been decomposed in the form
\begin{equation}\label{axpol}
\Pi_{\scriptscriptstyle A}^{\mu\nu}(k)
=\Pi_{\scriptscriptstyle A,T}(k^2)\left(g^{\mu\nu} 
- \frac{k^\mu k^\nu}{k^2}\right) 
+ \Pi_{\scriptscriptstyle A,L}(k^2)\frac{k^\mu k^\nu}{k^2}.
\end{equation}
Again, a bound axial-vector diquark corresponds to a pole in the $T$-matrix 
(\ref{tmaax}). The minimum values of the coupling ratios 
$g_{\scriptscriptstyle S}/g_\pi$ and $g_{\scriptscriptstyle A}/g_\pi$ 
required to get bound diquarks at zero density can be found in Table I.

Note that when the loop integrals are regulated with a simple cut-off on
either the 3- or 4-momentum, the longitudinal polarizability in this 
channel, $\Pi_{\scriptscriptstyle A,L}(k^2)$ does not vanish, in contrast to
the hybrid method involving dimensional regularization used by Ishii {\it
et al.}\cite{IBY95,Ish99}. However, since there is no conserved current coupled 
to these states, this does not violate any physical symmetries.

In the presence of quark matter, manifest Lorentz invariance is broken
and the structure of the polarization tensor 
$\Pi_{\scriptscriptstyle A}^{\mu\nu}(k)$ becomes more complicated. For a 
diquark momentum in the $z$-direction, $k=(k^0,0,0,k^3)$, it can be written
\begin{equation}\label{axpolm}
\Pi_{\scriptscriptstyle A}^{\mu\nu}(k^0,k^3)
=\left( \begin{array}{cccc}
A &\quad 0\quad &\quad 0\quad & Dk^0k^3 \\ 0 & B & 0 & 0 \\ 0 & 0 & B & 0 \\
Dk^0k^3 & 0 & 0 & B+C(k^3)^2  \end{array} \right).
\end{equation}
The use of a 3-momentum cut-off does lead to deviations from the Lorentz 
covariant structure shown in (\ref{axpol}) even in the vacuum case, but we have 
checked that these are small.

The results for diquarks at zero density are qualitatively similar to
those in Ref.~\cite{IBY95} despite a different choice of regulator and
the use of a nonzero current-quark mass in that work.  For a
constituent quark mass of $M=400$ MeV, as used in that work, we find
that the minimum value of $g_{\scriptscriptstyle S}/g_\pi$ for diquark
binding is 0.4 compared with 0.33 in Ref.~\cite{IBY95}. For
$g_{\scriptscriptstyle S}/g_\pi=0.6$ and $g_{\scriptscriptstyle
S}/g_\pi=0.8$ we get scalar diquark masses of 699 and 507 MeV
respectively, compared with 627 and 446 MeV (cf.~Table 2 of
Ref.~\cite{IBY95}). To bind the axial-vector diquark, we find a
minimum coupling strength $g_{\scriptscriptstyle A}/g_\pi=0.46$. The
corresponding value given in Ref.~\cite{IBY95} is 2.0, but this should
in fact be divided by a factor of 4 \cite{Ish99}, and so we again have
qualitative agreement with that work.

\section{THE FADDEEV EQUATION}
Because of the separability of the NJL interaction, the ladder approximation to
the Faddeev equation for the three-body system can be reduced to an effective 
two-body Bethe-Salpeter equation. This can be thought of as describing the 
interaction between a quark and a diquark, although it is not necessary that 
the diquark be bound. In our derivation of this equation, we have followed 
the procedure described in \cite{IBY95}. We reproduce here only the
main steps of this derivation; the details can be found in the original papers
\cite{IBY93,IBY95}. In the case of a purely scalar $qq$ interaction,
a thorough discussion of the Faddeev equation can also be found in 
Ref.~\cite{HT94}.  

\begin{figure}
\epsfysize=2.5cm \centerline{\epsffile{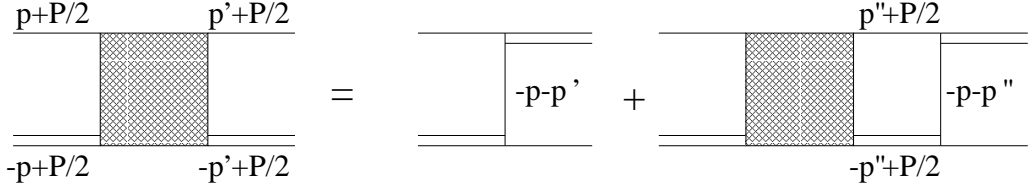}}
\caption{Diagrammatic representation of the Faddeev 
equation (\protect{\ref{eq:faddeev}}). \label{fadfig}}
\end{figure}

We denote the scattering amplitude of a quark on a diquark by 
$X_{ab}^{\alpha\beta}$. The indices $a,b$ label the diquark and, following
the convention of Ref.~\cite{IBY95} they take the values 5 (for the 
scalar diquark) and 0, 3, +1 and $-1$ (for the components of the axial-vector 
diquark in a spherical basis). These diquark indices will be written as 
subscripts or superscripts to indicate that the components of the axial diquark 
are covariant or contravariant respectively.
The Dirac indices $\alpha, \beta$ label the quark, taking the values 1 to 4. 
The amplitude obeys the integral equation, 
\begin{equation}\label{fad1}
X_{ab}^{\alpha\beta}(p',p) = Z_{ab}^{\alpha\beta}(p',p) 
+ \sum_{\gamma,c,\delta,d}\int \frac{d^4p''}{(2 \pi)^4} 
Z_{ac}^{\alpha\gamma}(p',p'')S^{\gamma\delta}(P/2+p'') 
\bar{\tau}^{cd}(P/2 - p'') X_{db}^{\delta\beta}(p'',p),
\label{eq:faddeev}
\end{equation}
corresponding to the diagram shown in Fig.~\ref{fadfig}. The piece of the kernel 
containing the propagator of the exchanged quark is
\begin{equation}
Z_{ab}^{\alpha\beta}(p',p)=\sum_{\gamma,\delta}\Omega_b^{\alpha\delta} 
S^{\gamma\delta}(-p-p')\bar{\Omega}_a^{\gamma\beta}, 
\end{equation}
where $\Omega$ and $\bar{\Omega}$ are the two-body vertex functions, already 
used in Eqs.~(\ref{kersc}) and (\ref{kerax}). The propagator of the spectator
quark is $S(P/2+p'')$, and $\bar{\tau}^{cd}(P/2 - p'')$ is the two-body
amplitude for the two interacting quarks. The separable nature of the two-body 
interaction means that this amplitude can be thought of as a diquark 
propagator, but one should remember that it describes the propagation of all 
two-quark states, not just the bound states (if they exist). It can be written 
as
\begin{equation}
\bar{\tau}^{cd} = \left( \begin{array}{cc} \bar{\tau}_{\scriptscriptstyle S} 
& 0 \\
 0 & \bar{\tau}_{\scriptscriptstyle A}^{\mu\mu'} \end{array} \right).
\end{equation}
where $\bar{\tau}_{\scriptscriptstyle S}$ and 
$\bar{\tau}_{\scriptscriptstyle A}^{\mu\mu'}$ are the scalar and 
axial-vector diquark ``propagators" given by Eqs.~(\ref{tmasca})
and (\ref{tmaax}) respectively.

To find bound states of the three-body system, we solve the homogeneous
version of equation (\ref{fad1}) for the effective two-body vertex function 
of a quark and a diquark. This vertex function, $X_{a}^{\alpha}$, is related to 
the amplitude $X_{ab}^{\alpha\beta}$ by 
\begin{equation}
X_{ab}^{\alpha\beta}(p',p) \rightarrow \frac{X_{a}^{\alpha}(p')
\bar{X}_{b}^{\beta}(p)}{P^2 - M_B^2 + i \epsilon}, \mbox{\hspace{1cm} as}
\qquad P^2 \rightarrow M_B^2,
\end{equation}
where $M_B$ is the mass of the bound baryon. From (\ref{fad1}) one obtains the 
equation
\begin{equation}\label{fad2}
X_{a}^{\alpha}(p) = \sum_{\gamma,c,\delta,d}\int \frac{d^4p'}{(2 \pi)^4} 
Z_{ac}^{\alpha\gamma}(p,p') S^{\gamma\delta}(P/2+p') 
\bar{\tau}^{cd}(P/2 - p') X_{d}^{\delta}(p'),
\end{equation}
for the vertex function.

This equation needs to be projected onto states of definite color, spin
and isospin. Projecting the kernel $Z_{ac}^{\alpha\gamma}$ onto a color-singlet
state gives
\begin{equation}\label{prker1}
Z_{ac}^{\alpha\gamma}(p,p') = -3 \left( \begin{array}{cc} \gamma_5 S(p+p')
\gamma_5 & \sqrt{3} \gamma_\mu S(p+p') \gamma_5   \\
\sqrt{3} \gamma_5 S(p+p') \gamma_{\mu'} & - \gamma_\mu S(p+p') \gamma_{\mu'}
\end{array} \right)_{\alpha \gamma},
\end{equation}
for the isospin-${1\over 2}$ channel, and
\begin{equation}\label{prker3}
Z_{ac}^{\alpha\gamma}(p,p') = -6 \Bigl(\gamma_\mu S(p+p') \gamma_{\mu'}
\Bigr)_{\alpha \gamma},
\end{equation}
for the isospin-${3\over 2}$ channel.

Next, a projection onto good spin and parity must be carried out. Like Ishii 
{\it et al.}\cite{IBY95} we use the helicity formalism of Jacob and 
Wick\cite{JW59}, constructing first a basis of states with definite helicity 
by acting with a rotation operator on helicity eigenstates whose momentum 
$\tilde{\bf p}$ lies along the $z$-axis: 
\begin{equation}\label{proj1b}
|{\bf p}(\omega),\alpha,a\rangle={\cal R}(\omega)|\tilde{\bf p},\alpha,
a\rangle
=\sum_{\alpha',a'} S^{\alpha' \alpha}(\omega)\hat{R}^{a'}_{\ \ a}(\omega)
|{\bf p}(\omega),\alpha', a'\rangle,
\end{equation}
where ${\bf p}(\omega)=R(\omega)\tilde{\bf p}$ lies in a general direction 
given by the Euler angles $\omega$.
Both the helicity $s_\alpha$ and the intrinsic parity $\eta_\alpha$ of the 
quark state are specified by the label $\alpha=1,$\dots,4, with $s_{1,3} = 
{1\over 2}$, $s_{2,4} = -{1\over 2}$, $\eta_{1,2}=+1$ and $\eta_{3,4}=-1$.
The helicity $\lambda_a$ of the diquark is specified by the label $a$, with 
$\lambda_{5,0,3}=0$, $\lambda_{+1}=+1$ and $\lambda_{-1}=-1$. The corresponding 
intrinsic parities are $\eta_{5,3,+1,-1}=+1$ and $\eta_0=-1$. Note that our
use of these labels differs slightly from that of Ishii {\it et 
al.}\cite{IBY95}. The rotation matrices appearing Eq.~(\ref{proj1b}) are
\begin{equation}\label{rotaq}
S^{\alpha' \alpha}(\omega) 
= \exp{(i s_{\alpha'} \psi)} \left( \begin{array}{cc}
d^{1/2}(\theta) & 0 \\ 0 &  d^{1/2}(\theta) \end{array} \right)_{s_{\alpha'}
s_\alpha} \exp{(i s_\alpha \phi)},
\end{equation}
and
\begin{equation}\label{rotadi}
\hat{R}^{a'}_{a}(\omega) = \exp{(i \lambda_{a'} \psi)} \left(
\begin{array}{ccc} 1 & 0 & 0 \\ 0 & 1 & 0 \\ 0 & 0 & d^1(\theta) \end{array}
\right)_{\lambda_{a'} \lambda_a}  \exp{(i \lambda_a \phi)},
\end{equation}
where the $d(\theta)$ are Wigner $d$-functions in Edmonds' 
convention\cite{Ed57}.

These basis states must now be projected onto good angular momentum:
\begin{equation}\label{angul}
|\bar p,\alpha,a; J M\rangle = \sqrt{\frac{2J+1}{8\pi^2}} 
\int d\omega\, {\cal D}^{J}_{M,s_\alpha + \lambda_a}(\omega) 
|{\bf p}(\omega),\alpha,a\rangle,
\end{equation}
where $\bar p=|{\bf p}|$. The 
resulting Faddeev equation for states of spin $J$ has the form
\begin{equation}\label{fad3b}
X^{\alpha}_{J,a}(p_0,p)=\sum_{\beta,b} \frac{1}{(2\pi)^5} \int
dp_0'\,{p'}^2dp'\, F^{\alpha \beta b}_{J,a}(p_0,p,p_0',p')
X^{\beta}_{J,b}(p_0',p').
\end{equation}
The expression for the kernel is given by the formula (D.9) of 
Ref.~\cite{IBY95}. We reproduce its form in our notation here, 
\begin{eqnarray}\label{kernelb}
F^{\alpha'\alpha a}_{J,a'}(p_0'\bar p',p_0 \bar p)
&=& -3 C^J_{a a'} (2 \pi)^2 \nonumber\\
&&\times\sum_{b,c} \int_0^\pi \sin\theta\,d\theta\, 
d^J_{s_\alpha + \lambda_a,s_{\alpha'} + \lambda_{a'}}(\theta) \nonumber \\ 
&&\qquad\qquad \times\displaystyle
\frac{1}{(p_0+p_0')^2-(\bar p^2+\bar p'^2+2\bar{p}\bar{p}' \cos\theta)
- M^2} \nonumber \\ 
&&\qquad\qquad \times\Bigl\{ \Bigl[ \gamma_{a'}^* \hat{d}^{1/2}(-\theta)\gamma^b 
p\llap/+{\tilde p\llap/}^\prime \gamma_{a'}^* \hat{d}^{1/2}(-\theta)\gamma^b 
\nonumber \\
&&\qquad\qquad\qquad\qquad - M \gamma_{a'}^* \hat{d}^{1/2}(-\theta)\gamma^b 
- 2 \hat{d}^{1/2}(-\theta)\hat{d}^{1}(-\theta)_{\lambda_{a'}\lambda_b} 
p\llap/ \nonumber \\
&&\qquad\qquad\qquad\qquad - 2 {\tilde p\llap/}^\prime \hat{d}^{1/2}(-\theta) 
\hat{d}^{1}(-\theta)_{\lambda_{a'}\lambda_b} + 2 M \hat{d}^{1/2}(-\theta) 
\hat{d}^{1}(-\theta)_{\lambda_{a'}\lambda_b} \nonumber \\ 
&&\qquad\qquad\qquad\qquad + 2 \hat{d}^{1/2}(-\theta) 
\hat{d}^{1}(-\theta)_{\lambda_{a'}\lambda_c} p_c\gamma^b \nonumber \\
&&\qquad\qquad\qquad\qquad + 2 \gamma_{a'}^* \hat{d}^{1/2}(-\theta) {\tilde 
p}^{\prime c} 
\hat{d}^{1}(-\theta)_{\lambda_{c}\lambda_b} \Bigr] 
S(P/2 + p) \Bigr\}_{\alpha'\alpha}\nonumber\\
&&\qquad\qquad\times\tau_b^a(P/2 -p).
\end{eqnarray}
In Eq.~(\ref{kernelb}), the asterisk denotes complex conjugation with respect 
to the explicit factor of $i$ in the spherical-basis components of vectors 
only. The matrices $\hat{d}^{1/2}(\theta)$ and $\hat{d}^1(\theta)$ are the ones 
appearing between the phase factors in Eqs.~(\ref{rotaq}) and (\ref{rotadi}) 
respectively. In the spin-${1\over 2}$ channel,
the numerical coefficients $-3C^{1/2}_{a'a}$ are the ones appearing in 
Eq.~(\ref{prker1}), with $C^{1/2}_{a'a}$ equal to +1 if both indices refer to 
scalar diquarks, $-1$ if both refer to axial diquarks, and $\sqrt 3$ if they 
are mixed.

Finally the Faddeev equation has to be projected onto positive parity states.
The resulting equation has the same form as (\ref{fad3b}), but with the 
positive-parity kernel,
\begin{equation}
F^{(+)\alpha\beta b}_{J,a} =F^{\alpha\beta b}_{J,a} + z_{\beta b}
F^{\alpha\bar\beta \bar b}_{J,a},
\end{equation}
where the quark index $\bar\beta$ is defined such that $s_{\bar\beta}=-s_\beta$ 
and $\eta_{\bar{\beta}}=\eta_\beta$, and the diquark index $\bar{b}$ is 
defined similarly. The phase factor is
\begin{equation}
z_{\beta b} = \eta_\beta \eta_b (-1)^{J-s-j_b},
\end{equation}
where $\eta_\beta$ and $\eta_b$ are the intrinsic parities of the quark and 
diquark, and their intrinsic spins are $s={1\over 2}$, $j_{5,0}=0$ and
$j_{3,+1,-1}=1$. The parity projection cuts the set of 20 coupled equations
down to 10. Of these, two describe only states with spin-${3\over 2}$ and so
decouple from the spin-${1\over 2}$ channel to leave 8 coupled equations.

A very similar set of equations can be derived for spin-${3\over 2}$ states. 
The kernel has the same form as given in Eq.~(\ref{kernelb}); only the
numerical coefficients differ from spin-${1\over 2}$ case. From 
Eq.~(\ref{prker3}) we see that $C^{3/2}_{a'a}$ is equal to $+2$ if both indices 
refer to axial diquarks and zero otherwise. Since the components involving 
scalar diquarks do not contribute, the number of coupled equations is again 
reduced to 8.

\section{NUMERICAL METHOD}
We solve the Faddeev equation (\ref{fad3b}) in the rest frame of the nucleon 
(or $\Delta$), where $P = (E,\vec{0})$. To avoid the singularities of the 
kernel, we perform a Wick rotation on the energy variables $p_0$ and
$p_0'$: $p_0  \rightarrow \alpha + i p_4 $ with 
\begin{equation}
\alpha = {M-m_{\rm diq}\over 2},
\end{equation}
where $m_{\rm diq}$ is the lower-energy pole in the diquark $T$-matrices,
(\ref{tmasca}) and (\ref{tmaax}) (i.e.~the mass of the lighter diquark).

Because of the Wick rotation, we have to solve 8 complex coupled integral
equations. We do so using the iterative method of Malfliet and Tjon 
\cite{MT69,RT88}. The set of equations may be written schematically as
\begin{equation}\label{fadschem}
K(E)\Phi=\Phi,
\end{equation}
where the kernel $K(E)$ depends nonlinearly on the energy eigenvalue $E$. 
Rather than solve this directly, we solve instead the linear eigenvalue 
problem
\begin{equation}
K(E)\Phi=\lambda(E)\Phi,
\end{equation}
for a fixed value of $E$. This is done iteratively, by acting with
$K(E)$ on some initial guess for the vector $\Phi$ of vertex functions
to obtain a new vector. This is repeated until the vectors of
functions in successive iterations are simply proportional to each
other. The proportionality constant is then equal to $\lambda(E)$. We search
on $E$ until we find a value for which $\lambda(E)=1$ and so the
solution $\Phi$ satisfies the original equation (\ref{fadschem}). In
the present problem, we use a simple initial guess consisting of a
Gaussian for the real part of each of the 8 components of $\Phi$, and
the derivative of a Gaussian for each of the imaginary parts. We find
that 4 or 5 iterations are generally enough to determine $\lambda(E)$.

The parameters of the model that describe the interactions in the $qq$ channels
are fixed using the solutions to the Faddeev equation at zero density. The 
values of the scalar and axial-vector couplings $g_{\scriptscriptstyle S}$ and 
$g_{\scriptscriptstyle A}$ that reproduce the masses of the nucleon and the 
$\Delta$ are listed in Table I for our two parameter sets.

Before describing our calculations at finite density, we should
compare our zero-density results with those in Ref.~\cite{IBY95}. Note
that apart from the difference in cut-off scheme and the use of
current quark masses, a further difference between our approach and
theirs is that Ishii {\it et al.} do not try to reproduce the nucleon
and $\Delta$ masses exactly, but investigate the range of parameters
$g_{\scriptscriptstyle S}/g_\pi$ and $g_{\scriptscriptstyle A}/g_\pi$
that can give bound states with reasonable masses. When only the
scalar coupling is included ($g_{\scriptscriptstyle A}/g_\pi=0$), we
find that the minimum value of $g_{\scriptscriptstyle S}/g_\pi$ to get
a bound nucleon is 0.8 compared with 0.5 in Ref.~\cite{IBY95}, both
for a quark mass of $M=400$ MeV.  Although these values are rather
different, one should note that the nucleon is very weakly bound in
the absence of the axial $qq$ coupling. For example, Ishii {\it et
al.} find a nucleon binding energy of only about 40 MeV for
$g_{\scriptscriptstyle S}/g_\pi=0.8$. It is therefore not too surprising
that the point at which the nucleon becomes bound is rather sensitive
to details, such as the choice of regulator or the use of the chiral
limit.

With both scalar and axial couplings the nucleon is more strongly bound and one 
might hope that results are less sensitive, but in this case it is harder to 
make meaningful comparisons with the results of Ref.~\cite{IBY95} because of 
the problem with the erroneous factor of 4 in the axial channel mentioned 
above. For a quark mass of $M=420$ MeV we are able to fit the nucleon and 
$\Delta$ masses with $g_{\scriptscriptstyle S}/g_\pi=0.735$ and 
$g_{\scriptscriptstyle A}/g_\pi=0.33$. The value for the axial-vector coupling 
is smaller than the minimum value to bind the $\Delta$ of 0.44 in
Ref.~\cite{IBY95}. The value for the scalar coupling is rather
somewhat larger than the range considered in that work. As in the case
of the diquarks, there are indications that our approach tends to give
less attraction in the scalar channel, compared with that of Ishii
{\it et al.}, but more attraction in the axial channel. Overall,
though, the results are qualitatively similar.

The Faddeev equation at finite density is solved using the same methods. In
this case the constituent quark mass is density dependent, and the
3-momentum of the valence quarks must be restricted to be larger than the Fermi 
momentum $k_F$. This acts as a lower cut-off of the momentum integrals:
$k_F < |{\bf p}_i| < \Lambda$ for the three-momenta ${\bf p}_i$ of all three
quarks. A similar Wick rotation is performed on the energy variables, but with
\begin{equation}
\alpha(k_F) ={\sqrt{M(k_F)^2 + k_F^2} - E_{\rm diq}(k_F)\over 2},
\end{equation}
where $E_{\rm diq}(k_F)$ is the lower-energy pole in the diquark $T$-matrices
at finite density.

\section{RESULTS}
\begin{figure} 
\epsfysize=12cm \centerline{\epsffile{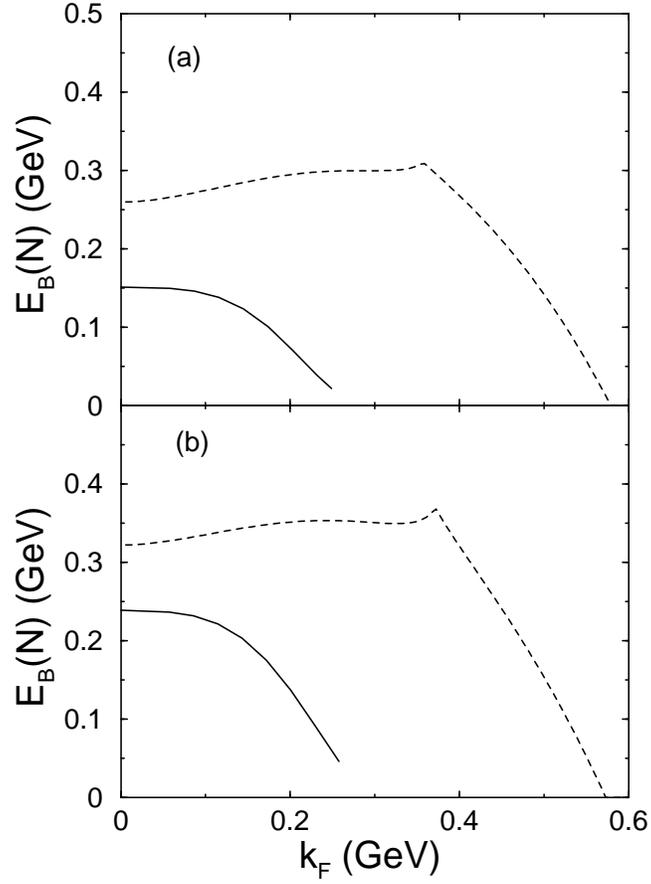}}
\caption{The binding energy of the nucleon (continuous curve) and of the
scalar diquark (dashed curve) as a function of $k_F$ for the parameter sets 
with (a) $M=450$ MeV and (b) $M=500$ MeV. \label{bindnuc}}
\end{figure}

We have solved the Faddeev equation at finite density for the energy $E_N$
of a nucleon at rest. In Figs.~\ref{bindnuc}(a) and \ref{bindnuc}(b) we
show the binding energy of the nucleon $B_N$ as a function of the Fermi 
momentum $k_F$ for our two sets of parameters. For these parameter sets the
scalar diquark is more strongly boundthan the axial, and so the nucleon binding 
energy is defined as
\begin{equation} \label{thresN}
B_N = E_S + \sqrt{k_F^2 + M^2} - E_N,
\end{equation}
with respect to the quark-diquark threshold. This is the relevant threshold
since the NJL model does not exhibit confinement. We compare $B_N$ to the
binding energy of the scalar diquark 
\begin{equation}
B_S = 2 \sqrt{k_F^2 + M^2} - E_S.
\end{equation} 
The binding with respect to the three-quark threshold is given by the sum of 
$B_N$ and $B_S$. One can see that the behaviors of the nucleon and
the scalar diquark at finite density are quite different. The binding energy of
the  diquark is very large and initially tends to increase with density, only 
decreasing when the Fermi momentum approaches the cut-off. In contrast, the 
binding of the nucleon decreases rather quickly, so that it is only marginally 
bound at nuclear matter density ($k_F=270$ MeV). 

\begin{figure}
\epsfysize=7cm \centerline{\epsffile{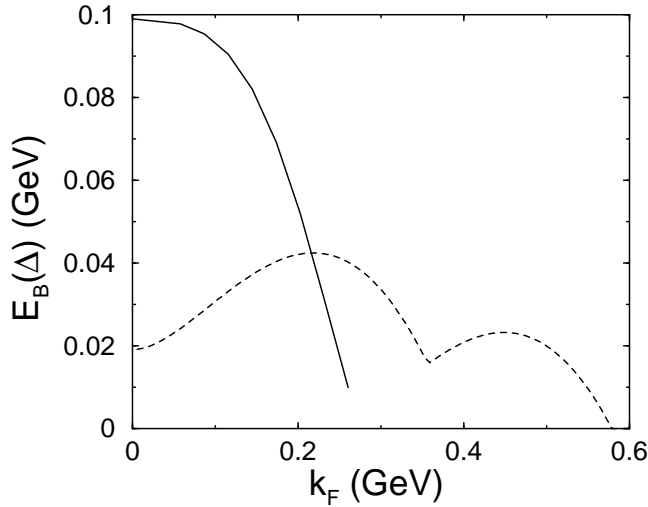}}
\caption{The binding energy of the $\Delta$ (continuous curve) and of the 
axial-vector diquark (dashed curved) as a function of $k_F$ for the parameter
set with $M=450$ MeV. \label{binddel}}
\end{figure}

We have also calculated the binding energy of the $\Delta$ as a function of 
$k_F$ and this is shown in Fig.~\ref{binddel}. In this case the threshold is 
determined by the binding energy of the axial diquark, which is also shown in
Fig.~\ref{binddel}. The behavior is similar to that found in the case of the 
nucleon. Since the axial coupling is smaller than the scalar coupling, the 
axial  diquark is less strongly bound than the scalar one. Nonetheless it
remains bound over the whole range of densities considered, while the $\Delta$ 
becomes unbound near nuclear matter density. Chiral symmetry restoration occurs 
at $k_F=359$ MeV and $k_F=372$ MeV for the parameter sets with $M=450$ MeV
and $M=500$ MeV respectively, which corresponds to the cusps in the diquark 
binding in Figs.~\ref{bindnuc} and \ref{binddel}.

\begin{figure}
\epsfysize=7cm \centerline{\epsffile{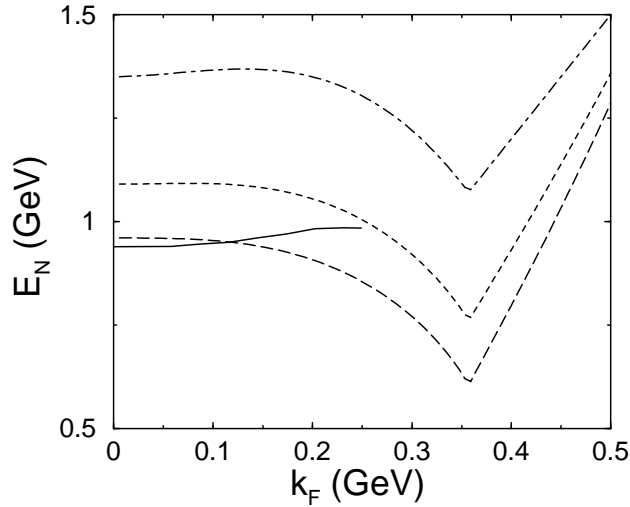}}
\caption{The energy of the nucleon (continuous curve) as a function of $k_F$ 
compared with $3/2$ times the mass of the scalar diquark (long-dashed curve),
the quark-diquark threshold (short-dashed curve) and the three-quark
threshold (dash-dotted curve), for the parameter set with $M=450$ MeV. 
\label{enernuc}}
\end{figure}

Another perspective can be gained by plotting the total energy of the
nucleon instead of its binding energy, as is done in Fig.~\ref{enernuc}.
For comparison, we have also plotted the quark-diquark threshold 
and the three-quark threshold. One can see that the nucleon energy
increases only slightly with density, while the quark-diquark threshold 
decreases more and more quickly as one approaches the density of chiral 
restoration (which is basically a consequence of the vanishing of the 
constituent quark mass at the transition). The density dependence of the
$\Delta$ energy shows the same qualitative behavior. 

To answer the question raised in the introduction about the possible 
competition between diquark and three-quark clustering, we have
also plotted the quantity ${3\over 2} E_S$. If $E_N < {3\over 2} E_S$
then a system of six quarks will prefer to form two nucleons, while if $E_N >
{3\over 2} E_S$ it will form three diquarks. Fig.~\ref{enernuc} shows that
nucleons are more stable than diquarks only for Fermi momenta smaller than 
about $130$ MeV, which corresponds to $1/8$ of nuclear matter density.
For our other parameter set, with $M=500$ MeV, nucleons remain more stable
than diquarks up to $k_F=170$ MeV, or $1/4$ of nuclear matter density. These
results clearly cast doubt on the validity of the model in this regime.

\section{CONCLUSIONS}     
We find that the NJL model predicts that, except well below nuclear
matter density, it is energetically much more favorable to form
three diquarks than it is to form two nucleons, and that there are no nucleon 
instabilities for the high densities where color superconductivity is expected
to occur.

However we should not conclude on the basis of these results that there is no 
competition between nucleon formation and diquark condensation. Since this
model predicts that nuclear matter should consist of diquarks, a result 
clearly at odds with what is actually observed, some elements of reality are 
missing. The most obvious of those is the lack of explicit confinement in the 
NJL model, something that is not easily remedied. It poses a rather deep 
question about the interpretation of the NJL model at finite densities. In much 
of the recent literature it has been used to model an instanton-induced
interaction between the quarks in high-density matter beyond the 
chiral/deconfining phase transition. We have tried to follow a more traditional 
approach, where the model is interpreted as a density model for hadron
structure in the vacuum. In that case the model contains {\it unphysical} 
diquark degrees of freedom, which may be ignored at zero density as being 
``irrelevant''. Both of these approaches have flaws. In the first 
(high-density) interpretation, we have no clue as to what remnants of 
confinement may play a role in the interaction. In the second (low-density) 
interpretation we cannot even describe nuclear matter properly.

Standing back and looking at our results in the light of these problems with
the model, we note from Fig.~\ref{enernuc} that the nucleon energy is roughly 
independent of density. It is mainly the increase in the diquark binding that 
renders the nucleon unstable. One could naively add to the NJL model a
three-body force that provides attraction in the color-singlet channel,
to incorporate an approximate description of confinement. Alternatively one 
might modify our treatment to use confined rather than free quark and diquark 
propagators \cite{Oet97}. Both of these choices have some appeal, but neither 
really deals with the underlying mechanism of confinement.

One should also remember that our results have all been obtained using
the ``rainbow-ladder'' approximation to the diquark Bethe-Salpeter 
equation and the Faddeev equation. In the context of models with nonlocal
interactions between the quarks, it has been shown \cite{BRS96,HAR97} that this 
approximation over-predicts diquark binding. With a toy model for the gluon
propagator which leads to quark confinement, diquark condensation has been
shown to occur even though diquarks are no longer bound at zero density 
\cite{BRS99}. Hence another way to approach this problem may be to study the 
Faddeev equation within a less restrictive calculational scheme. 

In conclusion, it is clear that the question of how confinement and the related 
three-quark correlations affect the phase structure of baryonic matter still 
remains to be answered.

\acknowledgments
We are very grateful to Noriyoshi Ishii for providing us valuable details
of the derivation of the Bethe-Salpeter and Faddeev equations. This work was 
supported by the EPSRC under grants GR/J95775 (Manchester) and GR/L22331 
(UMIST).

\end{document}